\documentclass[MSNbibl,nameyear,dvips]{arxstspdf}
\usepackage{flushend}
\usepackage{stfloats}

\volume{27}
\issue{3}
\pubyear{2012}
\firstpage{346}
\lastpage{347}
\doi{10.1214/12-STS381C} 
\referstodoi{10.1214/11-STS381}

\begin{document}
\begin{frontmatter}
\vspace*{6pt}
\title{An Answer to Multiple Problems with Analysis of Data on Harms?}
\runtitle{Discussion}

\begin{aug}
\author[a]{\fnms{Stephen} \snm{Evans}\corref{}\ead[label=e1]{stephen.evans@lshtm.ac.uk}}
\runauthor{S. Evans}

\affiliation{London School of Hygiene \& Tropical Medicine}

\address[a]{Stephen Evans is Professor, Department of Medical Statistics,
London School of Hygiene \& Tropical Medicine, Keppel Street, London, WC1E 7HT, UK \printead{e1}.}

\end{aug}



\end{frontmatter}


Randomized Controlled Trials have had many statistical developments
since their introduction into modern medicine sixty years ago. While
many of the advances have addressed general design and analysis \mbox{issues},
most have been motivated by or focused on assessment of efficacy in
contrast to safety. The reporting of harms is demonstrably weak
(Ioannidis and Lau, \citeyear{r5}; Loke and Derry, \citeyear{r7}) and, in spite of the
CONSORT guideline on harms (Ioannidis et al., \citeyear{r6}), continues to show
some deficiencies (Pitrou et al., \citeyear{r8}). Analytical developments focused
on harms have been very limited, reflecting lack of statistical effort
employed in that area as well as perhaps a general neglect of the less
exciting area of safety.

DuMouchel made a major contribution to extracting useful information
from spontaneous reports using Bayesian methods (DuMouchel, \citeyear{r2};
DuMou\-chel and Pregibon, \citeyear{r4}). This paper (DuMouchel, \citeyear{r3})
is potentially an important advance in assessing data on harms from
randomized trials. As an incidental point, it is possible it will also
have application in observational studies as well.

There are several really important features of\break MBLR as set out by
DuMouchel:

\begin{longlist}
\item[(1)] It addresses a clinically relevant problem, not addressed by
standard methods. The problem being that it is difficult if not
impossible to prespecify possible harms in terms of formal hypotheses
and the multiple medically related issues need to be seen as a
broad
picture as well as being reflected by narrow medical terms.\vadjust{\goodbreak} In practice,
also the data may be very limited because serious harms are rare for
medicines reaching the market.

\item[(2)] It addresses at least part of the problem of multiplicity of
many possible hypotheses of harm.

\item[(3)]  It avoids the epidemiological dilemma of lumping or splitting
terms which can lead to reduced sensitivity or simple loss of
statistical power. It also avoids the difficulty caused by composite
outcomes which, although they have their place in assessing efficacy,
have problems in that context but potentially worse problems in the
context of safety.

\item[(4)] It provides medically useful and interpretable estimates of
effects while retaining a good statistical foundation. The modeling is
consistent for each response variable related to a possible harm, and
can be used in trials which are primarily aimed at testing for efficacy.

\item[(5)] It does not seem heavily reliant on the particular form of the
Bayesian Priors being used.
\end{longlist}

The potential of the method is therefore very considerable and seems
destined to be used by the pharmaceutical industry and may eventually be
encouraged by regulators if it is shown in practice to be useful,
applicable and reasonably easily implemented.

Nothing in life is perfect though! It does require prespecification of a
group of medically-related terms, expressed as simple binary responses
and expected to behave in a similar manner (on a relative or odds ratio
scale)---showing exchangeability in Bayesian parlance. This may not
always be simple to do in practice, and even with the use of a
hierarchical medical dictionary like MedDRA which has over 16,000
``preferred terms,'' the choice of the number of terms and how wide a
range is included will not always be easy. There is then a danger that a
nonprespecified analysis may reach the conclusion desired by the analyst
or sponsor. It will be interesting to see if the MBLR method can be
applied to the more limited number of terms in the internationally
agreed ``Standardized MedDRA Queries'' which are already groupings of
terms from one or more MedDRA System Organ Classes (SOCs) that relate to\vadjust{\goodbreak}
a defined medical condition or area of interest. Similarly, it may be
possible to use a form of cluster analysis to provide medically sensible
groupings based on one set of data (not necessarily from RCTs) and then
to apply MBLR using these empirically derived groupings.

The covariates in this formulation also have to be expressed as binary
explanatory variables, though this is not a major problem.

The important idea of ``borrowing strength'' seems here to be less
dependent on the form of the prior than is the method suggested by Berry
and Berry (\citeyear{r1}); see Prieto-Merino and Evans (\citeyear{r9}).

One issue that may be of relevance is how well this borrowing of
strength works in slightly different contexts. In this example, all ten
terms are showing conventionally ``statistically significant''
differences with treatment so it could be argued that the MBLR adds
little to the understanding. Display 4 shows this and though there is
what may be a helpful effect for estimation of the OR for anuria, there
is (as expected given the data) relatively little effect elsewhere. The
consistent effect in this data set may be rather rare in other
situations where the MBLR might be used and more investigation of its
properties is required. The simulation provides some help and
reassurance that the method will perform well under some conditions, but
real life may be more complex.

The effects of sample size, number of covariates and number of response
variables studied on the overall results is not yet clear.

It is stated that the computational burden is not great and the
algorithms suggested do not require Gibbs sampling or Markov Chain Monte
Carlo methods, so making the method more accessible to medical
researchers. This reader is not sure how easy it will be to implement
MBLR in more standard software\vadjust{\goodbreak} and whether investigators will find it
easier than using WINBUGS or similar public domain software is not
clear.

In spite of these possible problems, the MBLR method shows very great
potential and is a real advance that will require further testing. It
may be a~way of doing meta-analysis of RCT data where multiple response
variables are studied both for efficacy and for safety. Might it avoid
the composite variable problems in analyses of RCTs for efficacy?
\vspace*{30pt}

\end{document}